\documentclass[twocolumn,showpacs,preprintnumbers,nofootinbib]{revtex4}
\usepackage{graphicx}% Include figure files
\usepackage{dcolumn}% Align table columns on decimal point
\usepackage{bm}% bold math
\usepackage{amssymb}  
\usepackage{amsmath}
\newcommand{\tr}{\mbox{tr}}
\def\beq{\begin{equation}}
\def\eeq{\end{equation}}
\def\bea{\begin{array}}
\def\eea{\end{array}}
\def\be{\begin{equation}}
\def\ee{\end{equation}}
\def\ba{\begin{eqnarray}}
\def\ea{\end{eqnarray}}

\def\to{\rightarrow}

\def\[{\left[}
\def\]{\right]}
\def\({\left(}
\def\){\right)}

\def\sm0{{\widetilde{m}_0}}

\def\U1em{{U(1)_{\rm em}}}
\def\to{\rightarrow}

\def\sq2{\sqrt{2}}

\def\ee{e^+e^-}

\def\End{\end{document}}
\def\Journal#1#2#3#4{{#1} {\bf #2}, #3 (#4)}
\def\NPB{{\rm Nucl. Phys.} B}
\def\PLB{{\rm Phys. Lett.}  B}
\def\PR{\rm Phys. Rep.}
\def\PRL{\rm Phys. Rev. Lett.}
\def\PRD{{\rm Phys. Rev.} D}

\def\fsl#1{\setbox0=\hbox{$#1$}                 % set a box for #1 
   \dimen0=\wd0                                 % and get its size
   \setbox1=\hbox{/} \dimen1=\wd1               % get size of /
   \ifdim\dimen0>\dimen1                        % #1 is bigger
      \rlap{\hbox to \dimen0{\hfil/\hfil}}      % so center / in box
      #1                                        % and print #1
   \else                                        % / is bigger
      \rlap{\hbox to \dimen1{\hfil$#1$\hfil}}   % so center #1
      /                                         % and print /
   \fi}  
\newcommand{\VEV}[1]{\langle #1 \rangle}

\begin{document}                                                              
%\draft
%@twocolumnfalse\endcsname

\title{New mechanism for the top-bottom mass hierarchy}%
\author{%
{\sc Michio Hashimoto\,$^1$,~~and  Shinya Kanemura\,$^2$}
}
\affiliation{%
%\address{\vspace*{5mm}
\vspace*{2mm} 
$^1$Department of Physics 
Pusan National University, Pusan 609-735, Korea\\
$^2$Department of Physics, 
Osaka University, Toyonaka, Osaka 560-0043, Japan
}
%\maketitle

\begin{abstract}
%\hspace*{-0.35cm}
We propose a mechanism to generate hierarchy 
between masses of the top and bottom quarks without fine tuning 
of the Yukawa coupling constants
in the context of the two Higgs doublet model (THDM).
In the THDM with a discrete symmetry, there exists the 
vacuum where only the top quark receives the mass of the order of
the electroweak symmetry breaking scale $v(\simeq 246\,\mbox{GeV})$, 
while the bottom quark remains massless. 
By introducing a small soft-breaking parameter $m_3^2$ of 
the discrete symmetry, 
the bottom quark perturbatively acquires a nonzero mass.
We show a model in which the small $m_3^2 [\sim v^2/(4\pi)^2]$ 
is generated by the dynamics above the cutoff scale of the THDM.
The ratio $\tan\beta$ of the two vacuum expectation values
is necessarily very large; i.e.,
$\tan \beta \sim m_t/m_b$. 
We also find a salient relation,
$1/\tan\beta \simeq m_3^2/m_H^2$, where
$m_H^{}$ is the mass of the extra CP-even Higgs boson.
Our scenario yields some specific features that can be tested 
in future collider experiments.
\pacs{12.60.-i,12.60.Fr,14.80.Cp \hfill   [\, \today \,]}

\end{abstract}

\maketitle

\setcounter{footnote}{0}
\renewcommand{\thefootnote}{\arabic{footnote}}

\section{Introduction} 

%{\it Introduction.}--- 
The measured quark mass spectrum shows a specific feature. 
Only the top quark has the mass of the order of the 
electroweak symmetry breaking (EWSB) scale $v$ 
($= (\sqrt{2} G_F^{})^{-1/2} \simeq 246$ GeV), 
while masses of the 
other quarks are much smaller. 
The top quark mass is $174$ GeV ($\simeq v/\sqrt{2}$),
while the bottom quark, the second heaviest, has the mass 
of $4.2$ GeV ($\ll v$).\cite{pdg}
In the Standard Model (SM),
however, the unique Higgs doublet field $\Phi_{\rm SM}^{}$ is
responsible for the EWSB and gives masses of 
all quarks via the Yukawa interactions; i.e.,   
$m_f \simeq y_f \langle \Phi_{\rm SM}^{} \rangle$ with
$\langle \Phi_{\rm SM}^{} \rangle = (0, v/\sqrt{2})^T \!$.
Therefore, the observed mass spectrum is obtained only 
by assuming unnatural hierarchy among the Yukawa coupling constants $y_f$. 
For instance, the hierarchy $y_b/y_t \simeq 1/40$ must be required for 
the top and bottom quarks. 
Nevertheless, no explanation for such fine tuning is given in the SM. 
%\Indent

In this paper, we propose an alternative scenario in which 
the quark mass spectrum is reproduced without fine tuning 
in magnitude of the Yukawa coupling constants. 
We study the hierarchy between $m_t$ and $m_b$ 
under the assumption of $y_t \sim y_b \sim {\cal O}(1)$. 
In order to realize $m_b/m_t \sim 1/40$ in a natural way,  
we consider the two Higgs doublet model (THDM) with 
$\Phi_1$ and $\Phi_2$, imposing the discrete $Z_2$ symmetry\cite{z2} 
under the transformation 
\begin{equation}
\Phi_1 \to - \Phi_1, \quad \Phi_2 \to + \Phi_2
\end{equation}
as well as 
\begin{equation}
 \left(\begin{array}{@{}c@{}}
  t \\ b
  \end{array}\right)_L \to 
+\left(\begin{array}{@{}c@{}}
  t \\ b
  \end{array}\right)_L , \quad 
  t_R \to + t_R, \quad b_R \to - b_R .
\end{equation}
Due to the $Z_2$ symmetry, only $\Phi_1$ couples to the bottom quark while 
$\Phi_2$ does to the top quark.
The hierarchy $m_t \gg m_b$ is then equivalent to
$v_2 \gg v_1$, where 
$\langle \Phi_{1,2} \rangle = (0, v_{1,2}/\sqrt{2})^T$.
We note that
there exists the vacuum with
$v_1 = 0$ and $v_2 = v$ when the $Z_2$ symmetry is exact. 
A nonzero value of $v_1$ $(\ll v_2)$ 
is induced as a perturbation of a small soft-breaking parameter 
$m_3^2$ for the $Z_2$ symmetry.
The small $m_3^2 [\sim v^2/(4\pi)^2]$ is generated by
the dynamics above the cutoff scale of the THDM.
We find a salient relation,
$1/\tan\beta \equiv v_1/v_2 \simeq m_3^2/m_H^2 \ll 1$, where
$m_H^{}$ is the mass of the extra CP-even Higgs boson.
Consequently, we obtain $m_b/m_t \ll 1$.
This scenario is extended to include the first two generation 
quarks. 
%\indent

We find that the extra Higgs bosons almost decouple with the 
weak gauge bosons in our model.
Moreover, the extra Higgs bosons as well as the SM-like one
turn out to have masses of the order of $v$.
The Higgs bosons with such masses are expected 
to be discovered at the CERN LHC because of 
the large value of $\tan\beta$\cite{lhc}. 
The characteristics of our scenario can further be tested 
by precision measurement at future linear colliders (LC's)\cite{lc}.

\section{Minimal Model} 
%{\it The Model.}--- 
The Lagrangian of the THDM with the 
softly-broken $Z_2$ symmetry is described as 
\begin{eqnarray}
{\cal L} &=& {\cal L}_{\rm kin}+{\cal L}_Y^{}-V, 
\end{eqnarray}
where ${\cal L}_{\rm kin}$ and ${\cal L}_Y^{}$ are the kinetic 
and Yukawa interaction terms, respectively. The 
Higgs potential $V$ is given by

\vspace*{-2mm}
\noindent
\begin{eqnarray}
V & = & m_1^2 |\Phi_1|^2 + m_2^2 |\Phi_2|^2 
 - \left[ m_3^2 \Phi_2^\dagger \Phi_1 + \mbox{(h.c.)}\right] 
 \nonumber \\ && 
 + \lambda_1 |\Phi_1|^4
 + \lambda_2 |\Phi_2|^4
 + 2\lambda_3 |\Phi_1|^2|\Phi_2|^2 \nonumber\\
&& + 2 \lambda_4 |\Phi_1^\dagger \Phi_2|^2  
 + \left[\, \lambda_5  \left(\Phi_2^\dagger \Phi_1\right)^2
   + \mbox{(h.c.)}\,\right],
\label{genTHDM}
\end{eqnarray}

\vspace*{-2mm}
\noindent
where 
$m_1^2$, $m_2^2$ and $\lambda_1$ to $\lambda_4$ 
are real, while $m_3^2$ and  $\lambda_5$
are complex. 
The Higgs doublet fields 
$\Phi_i$ $(i=1,2)$ with hypercharge $Y=1/2$ 
are parameterized by
% \vspace*{-2mm}
% \noindent
\begin{eqnarray}
\Phi_i = 
\left[
\begin{array}{c}
 \phi^+_i \\
 \frac{1}{\sqrt{2}} (v_i + h_i + i a_i)
\end{array}
\right],     
\end{eqnarray}
% \vspace*{-2mm}
% \noindent
where the vacuum expectation values (VEV's) $v_i$ $(i=1,2)$ 
satisfy $v_1^2+v_2^2=v^2$. 
The mass matrices for the Higgs bosons are diagonalized 
by mixing angles $\alpha$ and $\beta$.\cite{hhg} We then 
obtain five physical scalar states, $h$ and $H$ (CP-even), 
$A$ (CP-odd), and $H^\pm$ (charged), as well as 
three Nambu-Goldstone (NG) bosons, $\phi^0$ and $\phi^\pm$.
%\indent

We consider only the top and bottom quarks 
among fermions at first.
We discuss the extension for the other quarks later on.
In order to describe the assumption of $y_t \simeq y_b$, we 
introduce the global $SU(2)_R^{}$ symmetry\cite{cs,gs_2hdm}, 
in addition to the $SU(2)_L$ gauge symmetry:
% \vspace*{-2mm}
% \noindent
\begin{eqnarray}
  q_{L,R}^{} & \to & q_{L,R}' = U_{L,R}^{} \; q_{L,R}^{}\, , \\
  M_{21}^{}  & \to & M_{21}' = U_L M_{21} U_R^\dagger, 
\end{eqnarray}
% \vspace*{-2mm}
% \noindent
where $q_{L,R}^{} \equiv (t_{L,R}^{},b_{L,R}^{})$ and 
$U_{L,R} \in SU(2)_{L,R}$, respectively.
The $2 \times 2$ matrix $M_{21}^{}$ is defined by 
% \vspace*{-2mm}
% \noindent
\begin{eqnarray}
M_{21} \equiv \left( \tilde{\Phi}_2, \Phi_1 \right), \;\;\;
{\rm with} \;\;\;
\tilde{\Phi}_2 = i\tau_2 \Phi_2^\ast. 
\end{eqnarray}
% \vspace*{-2mm}
% \noindent
The $Z_2$ symmetry can be expressed in terms of $q_{L,R}^{}$ 
and $M_{21}^{}$ by
% \vspace*{-2mm}
% \noindent
\begin{eqnarray}
&&  q_L        \to  q'_L = q_L , \;\;\;
  q_R \to  q'_R = \tau_3 q_R , \\
&&  M_{21}^{}  \to  M_{21}' = M_{21} \tau_3.
\end{eqnarray}  
% \vspace*{-2mm}
% \noindent
The Yukawa interaction then is written as
\begin{eqnarray}
{\cal L}_Y^{} =
- y \bar{q}_L^{} M_{21}^{} q_R^{} 
 + \mbox{(h.c.)}\, , \label{L_Y}
\end{eqnarray}
% \vspace*{-2mm}
% \noindent
with $y \equiv y_t = y_b$. 
We also set 
\begin{equation}
 \lambda_1=\lambda_2=\lambda_3 (\equiv \lambda) \label{lam}
\end{equation}
in Eq.~(\ref{genTHDM}) to realize the $SU(2)_R^{}$ symmetry in 
quartic interactions.
The Higgs potential then is expressed by 
% \vspace*{-2mm}
% \noindent
\begin{eqnarray}
V(M_{21}^{}) &=& 
 \frac{1}{2}m^2 \tr(M_{21}^\dagger M_{21})
       -\frac{1}{2} \Delta_{12}\tr(M_{21}^\dagger M_{21}\tau_3)
       \nonumber \\ && 
\hspace*{-1.5cm} 
      - \left[ m_3^2 \,\det M_{21} + \mbox{(h.c.)}\,\right]
       + \lambda\left[\,\tr(M_{21}^\dagger M_{21})\,\right]^2
       \nonumber \\ && 
\hspace*{-1.5cm} 
       + 2\lambda_4 \det (M_{21}^\dagger M_{21})
       + \left[\, \lambda_5 (\det M_{21})^2+ \mbox{(h.c.)}\,\right], 
       \label{pot_M} 
\end{eqnarray}
% \vspace*{-2mm}
% \noindent
where 
$m^2 = m_1^2+m_2^2$ and  
$\Delta_{12} = m_1^2-m_2^2$. 
The $Z_2$ symmetry is softly broken 
by the mass term of $m_3^2$.  
A non-zero value of $\Delta_{12}$
measures the soft breaking of the global $SU(2)_R^{}$ symmetry. 
In order to evade explicit CP violation,
we choose the phases in $m_3^2$ and $\lambda_5$ 
to be zero. 
%\indent

We have introduced the global $SU(2)_R^{}$ symmetry only for the description
of $y_t = y_b$ in terms of a symmetry.
The Higgs potential also becomes simple since this symmetry requires 
the relation (\ref{lam}).
Our main results, however,  turn out to be unchanged even when this relation
is relaxed to some extent.
Cases without $SU(2)_R$ as well as those with CP violation will be discussed
in details elsewhere~\cite{hk}.

Let us consider the effective potential $V(\VEV{M_{21}})$
to study the vacuum structure. 
By using $SU(2)_L$ and $U(1)_Y$, 
the VEV's in the THDM can be generally 
parameterized as 
% \vspace*{-2mm}
% \noindent
\begin{eqnarray}
 \langle M_{21} \rangle =  \frac{1}{\sqrt{2}}
   \left( \begin{array}{cc} 
              v_2 & v_E^{}   \\
              0   & v_1 + i v_A^{} \\ \end{array}  \right). \label{M21}  
\end{eqnarray}
% \vspace*{-2mm}
% \noindent
Spontaneous breakdown of $U(1)_{\rm EM}$ and the CP symmetry occurs
if $v_E^{} \ne 0$ and $v_1 v_A^{} \ne 0$, respectively.  
We can easily show that 
the spontaneous $U(1)_{\rm EM}$ breaking cannot occur
at the tree level in our model.
The conditions for CP conservation are studied 
in Ref.~\cite{Gunion:2002zf}.
The effective potential is bounded from below
by the requirement of the vacuum stability\cite{2hdm_vs}, 
which leads to 
% \vspace*{-2mm}
% \noindent
\begin{equation}
  \lambda > 0, \quad 
  2 \lambda + \lambda_4 - |\lambda_5| >0.
\label{vs}
\end{equation}
% \vspace*{-2mm}
% \indent

We investigate details of the vacuum structure of our model 
in the tree level approximation. 
We first study the case with $m_3^2=0$ where 
the discrete $Z_2$ symmetry is exact.
We next include effects of $m_3^2 \ne 0$.

\begin{figure}[t]
\begin{center}
\resizebox{0.4\textwidth}{!}{\includegraphics{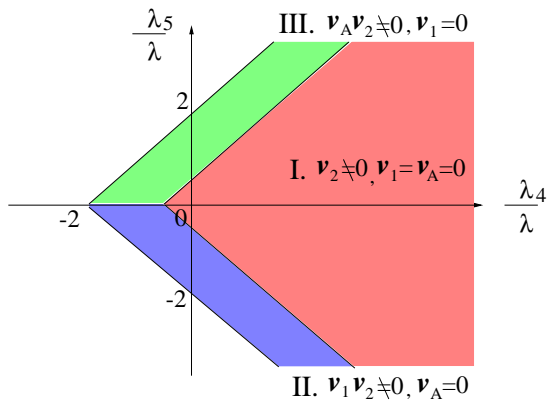}}
\end{center}
\vspace*{-2mm}
\caption{Vacuum structure for $m_3^2=0$ and $m_2^2 < -|m_1^2|$.}
\label{fig:phase}
\end{figure}

For $m_3^2 = 0$, the effective potential $V(\VEV{M_{21}})$ is
given by
\begin{eqnarray}
  V(\VEV{M_{21}}) &=& 
  \frac{m_1^2}{2} (v_1^2 + v_A^2) + \frac{m_2^2}{2} v_2^2
  +\frac{\lambda}{4} (v_1^2 + v_A^2 + v_2^2)  \nonumber \\ &&
  +\frac{\lambda_4}{2} (v_1^2+v_A^2)v_2^2
  +\frac{\lambda_5}{2} (v_1^2-v_A^2)v_2^2,
\end{eqnarray}
where we used Eq.~(\ref{M21}) with $v_E=0$.
The VEV's, $v_1$, $v_2$, and $v_A$, are determined by
the stationary conditions $\partial V(\VEV{M_{21}})/\partial v_i=0$, 
$(i=1,2,A)$.
Since spontaneous CP violation does not occur for $m_3^2 = 0$,
three types of the nontrivial vacuum are possible~\cite{2hdm_vs}:
\begin{description}
\item[(a)] $v_1 = v_A = 0, v_2 \ne 0$,
\item[(b)] $v_1 v_2\neq 0, v_A=0$,
\item[(c)] $v_A v_2\neq 0, v_1=0$.
\end{description}
In Fig.~\ref{fig:phase}, the area (I) corresponds to the vacuum (a), 
while the areas (II) and (III) do to the vacua (b) and (c), respectively.
Due to the vacuum stability conditions (\ref{vs}), 
there does not exist the stable vacuum out of the three areas.
Performing the transformation $\Phi_1 \to e^{i\pi/2}\Phi_1$ to 
the nontrivial vacuum (b), we obtain the vacuum (c).
The transformation corresponds to $\lambda_5 \to -\lambda_5$ 
in the Higgs potential with $m_3^2=0$.
The area (III) is thus the mirror image of the area (II).
%\indent

In order to realize $m_b/m_t \ll 1$ without fine tuning of Yukawa couplings, 
we choose the vacuum (a) which leads to
%\vspace*{-2mm}
%\noindent
\begin{eqnarray}
 m_t = \frac{1}{\sqrt{2}}\, y \,v,  \quad m_b = 0, 
\end{eqnarray} 
%\vspace*{-2mm}
%\noindent
because of $v_2=v$.
Although the bottom quark may receive a small mass
even in the vacuum (b),
the parameters of the Higgs potential must be very close to 
the boundary between the areas of (I) and (II).
This is fine tuning in a sense, so that we avoid such a case.
The vacuum (a) for $m_3^2=0$ is realized when\footnote{
There are the three vacua for $m_2^2 < -|m_1^2|$ 
as depicted in 
Fig.~\ref{fig:phase},
while the vacua (b) and (c) are squeezed out for the region (\ref{v1=0}).}
 
% \vspace*{-2mm}
% \noindent
\begin{eqnarray}
\!\!\!\!
\!\!\!\!
\!\!\!\!
&&m_2^2 < - |m_1^2|, \quad 
-\frac{\lambda_4}{\lambda}-\frac{\Delta_{12}}{-m_2^2} <
\frac{\lambda_5}{\lambda} < 
\frac{\lambda_4}{\lambda}+\frac{\Delta_{12}}{-m_2^2},
\label{v2 not 0} \\
\!\!\!\!
\!\!\!\!
\!\!\!\!
&& {\rm or} \hspace{18pt}
m_1^2 \geq -m_2^2 > 0 . \label{v1=0} 
\end{eqnarray}

% \vspace*{-1mm}
% \indent
Only the doublet $\Phi_2$ is responsible for the EWSB in the vacuum (a).
The doublet fields $\Phi_1$ and $\Phi_2$ do not mix
for $m_3^2=0$ 
because of the remaining $Z_2$ symmetry after the EWSB, 
$\Phi_1 \to -\Phi_1$.
The mass formulae of the physical Higgs bosons are 
% \vspace*{-2mm}
% \noindent
\begin{eqnarray}
  m_{h}^2 &=& 2 \lambda v^2,  \label{mh} \\
  m_{H^\pm}^2 &=& \Delta_{12}, \label{mHpm} \\
  m_{H}^2 
  &=& \Delta_{12} + (\lambda_4+\lambda_5) v^2, \label{mH} \\
  m_{A}^2 
  &=& \Delta_{12} + (\lambda_4-\lambda_5) v^2. \label{mA}
\end{eqnarray}
When $\Delta_{12}=0$, the charged Higgs bosons become 
the extra NG bosons associated with the breaking of 
the exact $SU(2)_R$ symmetry.
%\indent

We now switch on a {\it small} soft-breaking parameter $m_3^2 (\ll v^2)$ 
of the discrete $Z_2$ symmetry.
We do not consider the possibility of spontaneous CP violation~\footnote{
This subject will be addressed in Ref.~\cite{hk}.}.
A nonzero $v_1$ is necessarily induced for $m_3^2 \ne 0$ from 
the stationary condition.
As a perturbation from the vacuum (a) with $m_3^2 = 0$,
we consequently obtain 
% \vspace*{-2mm}
% \noindent
\begin{eqnarray}
   \frac{v_1}{v_2}  (\equiv \frac{1}{\tan\beta})  
          &=& \frac{m_3^2}{m_H^2} 
                 \left\{ 
     1 +{\cal O}\left(\frac{m_3^4}{v^4}\right) \right\},  
\label{ratio}
\end{eqnarray}
% \vspace*{-2mm}
% \noindent
where we used the tree-level mass formula in Eq.~(\ref{mH}).
Because of $v_1^2+v_2^2 = v^2$, 
the expression for $v_2$ is slightly modified to
$v_2 = v [1-{\cal O}(m_3^4/v^4)]$ from $v_2=v$.
The masses of the top and bottom quarks are given by 
% \vspace*{-2mm}
% \noindent
\begin{eqnarray}
 m_t \simeq \frac{1}{\sqrt{2}}\, y \,v, \quad
 m_b = \frac{1}{\sqrt{2}}\, y \,v_1, \label{mt-mb}
\end{eqnarray} 
% \vspace*{-2mm}
% \noindent
so that the bottom quark finally obtain the small mass. 
The mass hierarchy of $m_t$ and $m_b$ then is deduced from Eqs.~(\ref{ratio})
and (\ref{mt-mb}) without fine tuning of the Yukawa coupling constants; i.e.,
$m_t/m_b = \tan \beta$.
With nonzero $m_3^2$ the Higgs doublets $\Phi_1$ and $\Phi_2$
do mix. The mixing angle $\beta-\alpha$ is expressed as
% \vspace*{-2mm}
% \noindent
\begin{equation}
 \sin (\beta-\alpha) \!= \! 1 
 - \left(\frac{m_H^2-m_{H^\pm}^2}{m_H^2-m_h^2}\right)^2
     \!\! \frac{2}{\tan^2 \beta}
 + {\cal O}\left(\frac{m_3^6}{v^6}\right), \label{sin-b-a}
\end{equation}
% \vspace*{-0mm}
% \noindent
where Eqs.~(\ref{mh})--(\ref{mA}) and $\tan \beta \gg 1$ are used.
From Eqs.~(\ref{mh}) and (\ref{sin-b-a}), the property of 
the CP-even Higgs $h$ is similar to the SM one.
We note that Higgs boson masses in 
Eqs.~(\ref{mh})--(\ref{mA}) receive corrections of 
${\cal O}(m_3^4/v^4)$. 
These 
corrections, however, do not affect the expressions in 
Eqs.~(\ref{ratio}) and (\ref{sin-b-a}). 
%\indent

Let us estimate the typical size of the masses of the extra Higgs bosons.
The value of $\tan\beta$ is fixed by
$\tan\beta = m_t/m_b \sim 40$.
On the other hand, the small value of $m_3^2 (\ll v^2)$ can be interpreted
as $m_3^2 \simeq v^2/(4\pi)^2$.
In the next section, we shall present a concrete model in which 
such a small $m_3^2 [\simeq v^2/(4\pi)^2]$ is radiatively induced 
by the dynamics above the cutoff scale of the THDM. 
From Eq.~(\ref{ratio}), the mass of $H$ is expressed as 
\begin{equation}
 m_H^2 \simeq m_3^2 \tan\beta.  
\end{equation}
Therefore, the size of $m_H$
is at most of the order of $v$.
Furthermore, the masses of $A$ and $H^\pm$ are also the same order
because of the relations
\begin{equation}
 m_{H^\pm}^2 = m_H^2 - (\lambda_4+\lambda_5)v^2, \quad
 m_A^2 = m_H^2 - 2\lambda_5 v^2 , \label{m_gen}
\end{equation}
which are obtained from Eqs.~(\ref{mHpm})--(\ref{mA}).
%\indent

We have found Eqs.~(\ref{ratio}) and (\ref{sin-b-a}), 
assuming the softly broken $SU(2)_R$ symmetry; i.e., 
$\lambda_1=\lambda_2=\lambda_3(=\lambda)$.
We now give comments on the case with
$\lambda_1 \ne \lambda_2 \ne \lambda_3$, 
relaxing the $SU(2)_R$ symmetry.
First, it can be shown that Eq.~(\ref{ratio}) does not change.
Second, although Eq.~(\ref{sin-b-a}) is slightly modified, 
the essential result of $\sin (\beta-\alpha)=1-{\cal O}(\tan^{-2}\beta)$ 
still holds.
Finally, the masses of the extra Higgs bosons remain ${\cal O}(v)$
even for $\lambda_1 \ne \lambda_2 \ne \lambda_3$,
because Eq.~(\ref{m_gen}) turns out to be unchanged as well. 

\begin{figure}[t]
\begin{center}
 \begin{tabular}{cp{2mm}c}
  \resizebox{0.20\textwidth}{!}{\includegraphics{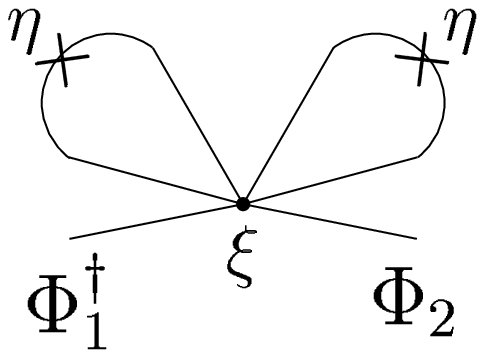}} &&
  \resizebox{0.18\textwidth}{!}{\includegraphics{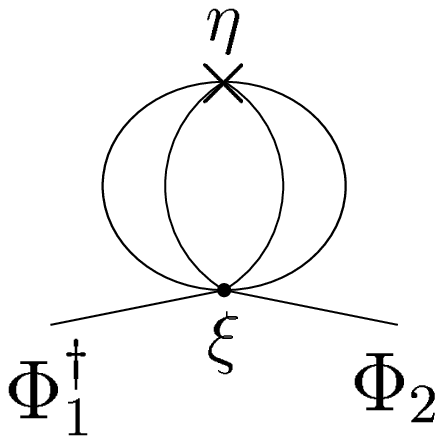}} \\
  (Case A) && (Case B)
 \end{tabular}
\end{center}
\vspace*{-5mm}
\caption{Feynman diagrams to induce a small $m_3^2$.}
\label{fig:m3}
\end{figure}

\section{A mechanism for small $m_3^2$}
%{\it A mechanism for small $m_3^2$.}--- 
We discuss an example where the small $m_3^2$ 
is generated radiatively in the low energy scale.
Let us consider a model with a complex scalar field 
$S$ 
which is a $SU(2)_L$ singlet without $U(1)_Y$ charge. 
The Lagrangian is given by 
% \vspace*{-2mm}
% \noindent
\begin{eqnarray} 
{\cal L} &=& {\cal L}_{\rm kin}
           - {V}_{\Phi} 
           - {V}_{S}  
           - {V}_{\fsl{Z}_2}, \label{lagm3}
\end{eqnarray} 
% \vspace*{-2mm}
% \noindent
where ${\cal L}_{\rm kin}$ represents the kinetic term and
$V_{\Phi}^{}$ is the $Z_2$ symmetric part\footnote{
We here concentrate on the mechanism to induce $m_3^2$, 
assuming that the $Z_2$ invariant part
$V_{\Phi}^{}$ comes from some other dynamics.
} 
of the THDM potential~(\ref{genTHDM}) with $m_3^2=0$. 
The potential $V_S$ for the complex scalar $S$ and the interaction term 
$V_{\fsl{Z}_2}$ between $S$ and $\Phi_{1,2}$ are given by  
% \vspace*{-3mm}
% \noindent
\begin{eqnarray} 
{V}_{S} =  M_S^2 S^\dagger S
            + \kappa (S^\dagger S)^2
            + {V}_{Z_{2n}}^{}, \;\; M_S^2 > 0, \label{VS}
\end{eqnarray} 
% \vspace*{-2mm}
% \noindent
with 
% \vspace*{-2mm}
% \noindent
\begin{eqnarray} 
 V_{Z_{2n}}^{} = \frac{\eta}{\Lambda^{2n-4}} 
         \left( S^{2n} + {\rm h.c.} \right), \;\; \eta \sim {\cal O}(1),
 \label{VM}
\end{eqnarray} 
% \vspace*{-2mm}
% \noindent
and 
% \vspace*{-2mm}
% \noindent
\begin{eqnarray} 
V_{\fsl{Z}_2}
            = \frac{\xi}{\Lambda^{2\ell-2}} 
             \left( S^{2\ell} \Phi_1^\dagger \Phi_2
            + {\rm h.c.} \right),\;\; \xi \sim {\cal O}(1), \label{VB}   
\end{eqnarray} 
% \vspace*{-2mm}
% \noindent
respectively. In Eqs.~(\ref{VM}) and (\ref{VB}), $\Lambda$ denotes
the cutoff scale of the model. 
We now set $n=1$ (case A) or 
$n=\ell$ (case B) with $\ell \geq 1$.
We note that $V_S$ has the $Z_{2n}$ symmetry under 
$S \to e^{i\frac{\pi}{n}} S$,
while $V_{\Phi}$ is $Z_2$ invariant under the transformation
$\Phi_1 \to - \Phi_1, \;\;\Phi_2 \to + \Phi_2$. 
The interaction term~(\ref{VB}) 
explicitly breaks both $Z_{2n}$ and $Z_2$. 
Some invariant terms under $Z_{2n}$ and $Z_2$ are not 
explicitly included here, as they are irrelevant 
to our conclusion. 
%\indent

Supposing that $M_S (\sim \Lambda)$ is much larger than the EWSB scale,
we integrate out the field $S$ and thereby obtain the THDM 
with the softly-broken $Z_2$ symmetry $(m_3^2 \ne 0)$
as the low-energy effective theory.
From the Feynman diagrams depicted in Fig.~\ref{fig:m3},
we estimate
%\vspace*{-4mm}
\begin{align}
& m_3^2 \sim \xi \eta^{\ell} \frac{1}{(4\pi)^{2\ell}} M_S^2, &
{\rm for \;\;\; Case \; A}, \\
& m_3^2 \sim \xi \eta \,\frac{1}{(4\pi)^{2(2\ell-1)}} M_S^2, &
{\rm for \;\;\; Case \; B}.
\end{align}
% \vspace*{-2mm}
% \noindent
For example,
we can obtain $m_3^2 \sim v^2/(4\pi)^2$ for $\ell=2$,
if we take the cutoff $M_S = 4\pi v$ for 
Case~A or $M_S = (4\pi)^2 v$ for Case~B.
For $\ell =2$, we do not need higher dimensional operators 
except for the $Z_2$ breaking term~$V_{\fsl{Z}_2}$.
%\indent

We may consider other possibilities to obtain small $m_3^2$ values
based on many ideas such as 
Topcolor instanton\cite{topc} and large extra dimensions\cite{exd}.
Also useful is a model which provides effectively 
$(\Phi_1^\dagger \Phi_2)^3$ with a coefficient $\sim {\cal O}(1)$ 
while prohibits the hard breaking terms of the $Z_2$ symmetry such as 
$(\Phi_1^\dagger \Phi_1) (\Phi_1^\dagger \Phi_2)$.

\section{Quark mass matrices}

%{\it Quark mass matrices.}---
We discuss the extension of our model 
incorporating first two generation quarks. 
Can we reproduce the observed quark mass spectrum 
and the Kobayashi-Maskawa (KM) matrix? 
%\indent

Under the discrete symmetry\cite{z2}, 
two types of Yukawa interactions are possible in the THDM, so called 
Model~I and Model II\cite{hhg}. 
The flavor changing neutral current (FCNC) then does not 
appear at the tree level\cite{z2}. 
Obviously Model I is inconsistent with our scenario, so that  
we here study Model II, 
%\vspace*{-2mm}
\begin{eqnarray}
 - {\cal L}_Y \!\!= \!\!\!\!\sum_{i,j=1}^3 \!\!
               \left(
                Y^{ij}_{D} \overline{q}_L^{(i)} \Phi_1 D_R^{(j)}
          \!+\! Y^{ij}_{U} \overline{q}_L^{(i)} \tilde{\Phi}_2 U_R^{(j)}
               \right)
          \!+\! {\rm (h.c.)} , \label{type2}
\end{eqnarray}
% \vspace*{-2mm}
% \noindent
where $q_L^{(i)}$ is the left-handed quark doublet of the $i$-th
generation, and $D^{(i)}_{R}=(d_{R},s_{R},b_{R})^T$ 
and $U^{(i)}_R=(u_{R},c_{R},t_{R})^T$. 
We then assume that matrices of the Yukawa coupling take the following forms,
%\vspace*{-1mm}
\begin{equation}
Y^{ij}_U \sim Y^{ij}_D \sim y \,
\left( 
\begin{array}{ccc} 1 & 1 & 1 \\ 1 & 1 & 1 \\ 1 & 1 & 1 \end{array}
\right), 
\quad y \sim {\cal O}(1),
\end{equation}
% \vspace*{-2mm}
% \noindent
which lead to $m_t \gg m_c,m_u$ and $m_b \gg m_s,m_d$, and
the KM matrix becomes approximately diagonal.  
We can numerically reproduce the data for the mass spectrum 
and the KM matrix\cite{pdg},
allowing fluctuations of the Yukawa coupling constants,
%\vspace*{-1mm}
\begin{eqnarray}
Y_{U}^{ij} = y\, \epsilon_{ij}^U, \;\;
Y_{D}^{ij} = y\, \epsilon_{ij}^D, \;\;
{\rm with} \;\; 0.5 < |\epsilon_{ij}^{U,D}| < 1.5. \label{ymat}
\end{eqnarray}
%\indent

Three comments are in order:
({\it a}) Although we can avoid hierarchy among Yukawa couplings,
subtle cancellation among the ${\cal O}(1)$ mass-matrix elements 
is required to obtain masses of light quarks.
({\it b}) We may adopt Model III~\cite{type3} 
to our scenario, if the FCNC is suppressed
by some mechanism.
({\it c}) It is possible to apply our scenario to the lepton sector. 
The $\tau$ lepton then receives the small mass
due to the similar mechanism to the bottom quark.
At the same time, however, 
the Dirac mass of the tau neutrino could be produced 
around $m_t$.
To explain the tiny (Majorana) mass of the tau neutrino, 
additional mechanism such as 
the Seesaw\cite{seesaw} might be helpful. 
%\indent

\section{Summary and Discussions}
%{\it Summary and Discussions.}---
We have proposed the mechanism to explain the mass hierarchy 
between the top and bottom quarks without fine tuning, 
starting from the vacuum with 
$(v_1,v_2)=(0,v)$.
Such a vacuum can exist when the $Z_2$ symmetry is exact.
The observed mass spectrum $m_t \gg m_b \ne 0$ is realized
via the small soft-breaking parameter $m_3^2$ for the $Z_2$ symmetry.
We have presented the model in which a small $m_3^2$ is
induced from the underlying physics above the cutoff scale of the
THDM. 
%\indent

The phenomenological implication is as follows. 
The size of $\tan\beta$ corresponds to the ratio $m_t/m_b \sim 40$.
We have found the relation 
$m_H^2 \simeq m_3^2 \tan\beta \sim {\cal O}(v^2)$.   
Therefore, the masses of the extra Higgs bosons $H$, $A$ and $H^\pm$ 
are expected to be ${\cal O}(v)$. 
The THDM with such parameters is constrained 
by the theoretical considerations\cite{2hdm_pu,2hdm_triv} as well as 
the available data.   
When $m_{H^\pm} \simeq m_H$, or $m_{H^\pm} \simeq m_A$, 
our model can satisfy the constraint from the LEP precision data\cite{pdg}.
The mass of the charged Higgs boson in our scenario 
may not conflict with the $b \to s \gamma$ result\cite{bsgamma}.
The doublet $\Phi_2$ is mainly responsible for the EWSB, 
so that we obtain $\sin(\beta-\alpha) \simeq 1$ in a good
approximation.
%\indent

In addition to the SM-like Higgs boson $h$, 
all the extra Higgs bosons in our model
are expected to be discovered at the LHC.
Our prediction of $\sin (\beta-\alpha) \simeq 1$ can also be confirmed
at the LHC and LC's.
Our scenario may further be tested by measuring 
the $hhh$ coupling at future LC's\cite{hhh}.
More detailed phenomenological analysis will 
be done elsewhere\cite{hk}.
%\indent

\section*{Acknowledgements}

The authors thank Yasuhiro Okada for useful comments. 

%\vspace*{-4mm}

\end{document}